\documentclass[%
twocolumn,
groupedaddress,
showpacs,
 amsmath,
 amssymb,
 prl,
]{revtex4-1}

\usepackage{color}
\usepackage{graphicx}
\usepackage{dcolumn}
\usepackage{bm}

\begin{document}
\everymath{\displaystyle}

\title{Resonant and antiresonant bouncing droplets}
\author{M. Hubert}
\author{D. Robert}
\author{H. Caps}
\author{S.Dorbolo}
\author{N. Vandewalle}
\affiliation{GRASP, Physics Dept., University of Li\`ege, B4000 Li\`ege, Belgium.}
\date{\today}
\begin{abstract}
When placed onto a vibrating liquid bath, a droplet may adopt a permanent bouncing behavior, depending on both the forcing frequency and the forcing amplitude. The relationship between the droplet deformations and the bouncing mechanism is studied experimentally and theoretically through an asymmetric and dissipative bouncing spring model. Antiresonance effects are evidenced. Experiments and theoretical predictions show that both resonance at specific frequencies and antiresonance at Rayleigh frequencies play crucial roles in the bouncing mechanism. In particular, we show that they can be exploited for droplet size selection. 
\end{abstract}
\maketitle 

Bouncing Droplets (BD) on vibrated liquid interfaces attract much attention because of their peculiar properties \citep{Couder2005}. An air layer separates the droplet from the vibrated surface which can therefore bounce vertically upon the liquid without coalescing. BD have the great advantage to transport some quantities of liquid without chemical contamination \citep{Eddi2008,Gier2012}. Moreover, BD may be either fragmented \citep{Gilet2007} or used to create controlled microemulsions \citep{Terwagne2010}. It is suggested that those droplets may be used in some microfluidics applications \citep{Vdw2009}. Beyond fluid mechanics, BD also reproduce experiments originally reserved for the quantum world. Indeed, due to the coupling between the droplet and the wave created on the liquid surface by successive impacts \citep{Protiere2006}, BD may start to walk horizontally along the liquid. Such droplet-wave association exhibits physical phenomena analogous to the ones encountered in quantum mechanics : tunnel effect \citep{Eddi2009}, diffraction \citep{Couder2012}, double-slit experiment \citep{Couder2006}, quantification of revolution orbits \citep{Fort2010} and Zeeman splitting of energy levels \citep{Eddi2012}. Resonance into cavities are also investigated \citep{Fort2013,Harris2013}. Depending on the forcing parameters, various bouncing modes are also observed, from simple bounces to period doubling and much more complex trajectories \citep{Terwagne2013,Willassen2013}. Nevetheless, those dynamics  highly depend on the droplet radius as illustrated in \citep{Protiere2006,Dorbolo2008}. Thus, reproductibility is a challenging problem. In order to understand such a rich dynamics, models have been proposed. While some of them focus on the surface deformation or on the effect of the air layer, only a few of them consider the possible droplet deformations. If they do, they also consider wave propagation or lubrification which make those models particularly complex \citep{Gilet2008,Willassen2013}. 

In this letter, we focus on BD deformations and their link with the bouncing dynamics without considering lubrification or wave propagation on the liquid surface. For this purpose, we investigate the BD dynamics experimentally and we propose a simple model consisting in an Asymmetric Bouncing Spring (ABS). Antiresonance is evidenced and rationalized. From the particular feature of the resonance and the antiresonance, we also propose a way to select bouncing droplet size.

\begin{figure}[!]
	\centering
	\includegraphics[scale=.25]{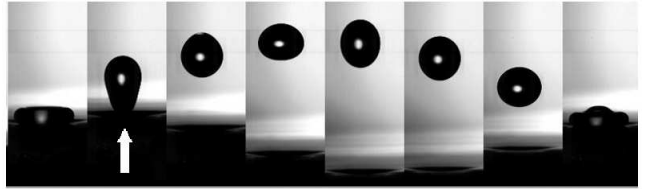} 
	\caption{BD for frequency $f=50$ Hz, radius $a=0.76$ mm and viscosity $\nu=5$ cSt. One observes that the droplet adopts periodically oblate and prolate shapes. Note that the deformation may be asymmetric when the droplet takes off as denoted with a white arrow.}{\label{fig:bounces}}
\end{figure}

\begin{table}
	\begin{center}
		\begin{tabular}{|c|c|c|c|}
			\hline Droplet viscosity $\nu$ (cSt) & 5 & 20 & 50\\
			\hline Surface tension $\sigma$ (N/m$^2$) & 1.97 10$^{-2}$ & 2.06 10$^{-2}$ & 2.08 10$^{-2}$  \\
			\hline Density $\rho$ (kg/m$^3$) & 910 & 949 & 960 \\
			\hline Dissipation $\xi$  &  0.250 & 0.257 & 0.324 \\
			\hline Mass distribution $\mu$  & 0.659 & 0.691 & 0.743 \\
			\hline
		\end{tabular}
	\end{center}
	\vskip -0.4cm
	\caption{Parameters relative to the silicon oils used in our experiments. The three first rows contain the fluid properties while the last two rows contain the fitted parameters from Eq.(\ref{eqs:Gamma}).}{\label{tab:parameter}}
\end{table}

The experimental setup consists in a container filled with highly viscous silicone oil (Dow Corning 200 Fluid, $\nu$=1000 cSt) in order to inhibit the surface deformations. Droplets of viscosity ranging from 5 to 50 cSt are formed with a needle. The information relative to the silicon oils are summed up in Table I. The container is vertically shaken with a pulsation $\omega$ and an amplitude $A$. The maximum acceleration normalized by gravity $\Gamma$=$A\omega^2/g$ is accurately measured by an accelerometer. Figure \ref{fig:bounces} presents snapshots of the bounce for a specific frequency. One observes deformations of the droplet which experiences a periodic change from oblate to prolate shapes. Moreover, when the droplet detaches from the interface on the crest of each oscillation, some asymmetry in the droplet shape is seen. This observation is pointed by a white arrow on Fig.\ref{fig:bounces}. Those observations suggest that droplet deformations superimpose with the periodic forcing from the surface. This hypothesis has been studied in earlier works \citep{Dorbolo2008,Biance2006}. \\

\begin{figure}[h!]
	\centering
	\includegraphics[width=9cm]{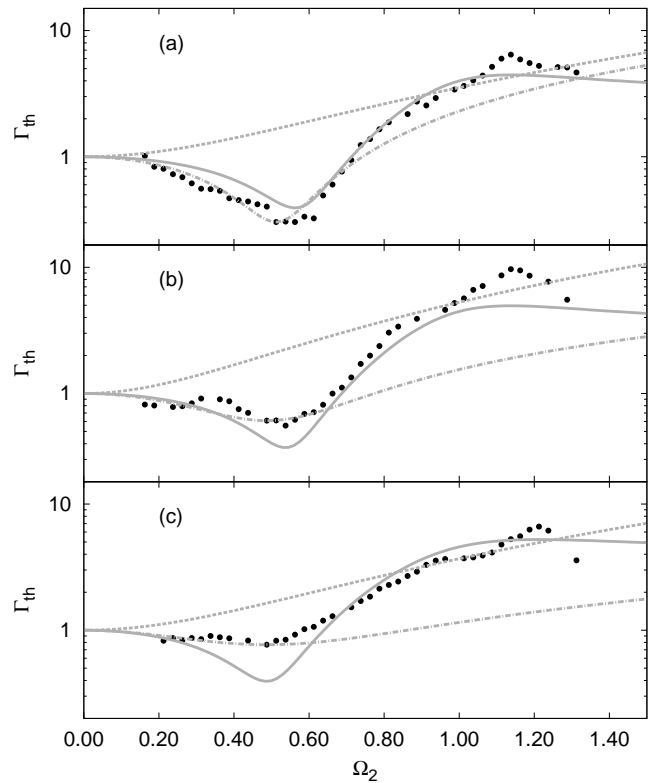}
	\caption{Bouncing threshold (logarithmic scale) as a function of the dimensionless Rayleigh frequency $\Omega_2$ for droplets of different viscosities: (a) 5cSt, (b) 20 cSt and (c) 50cSt . Black dots represents the exprimental data. The grey curves corresponds to anaytical models. Dotted: Couder model \citep{Couder2005}, Dashed: Eichwald model \citep{Eichwald2010}, Plain: ABS model. }{\label{fig:Comp}}
\end{figure}

Let us consider the plots of Fig.\ref{fig:Comp}. This figure presents the bouncing threshold $\Gamma_{th}$, in a logarithmic scale, as a function of the dimensionless frequency $\Omega_2$, as defined later in Eq.(\ref{eqs:Rayleigh}). By means of this rescaling, for each viscosity, droplets of different radii can be compared in a single curve. Those graphs show resonances at specific frequencies. For the resonant frequencies of Fig.\ref{fig:Comp}, the droplet bounces onto the surface for $\Gamma_{th}<1$; i.e. for maximal acceleration below $g$. Indeed, in order to overcome gravity, the droplet stores some elastic energy into its deformation and uses this energy for taking off. This observation points out the importance of deformation in the bouncing dynamics. Resonance in BD dynamics has already been studied. Further information can be found in \citep{Dorbolo2008}. Others features of those curves have however not been reported. One remarks that, for specific frequencies $\Omega_2 \approx 1$, the droplet bounces only for very high $\Gamma$ values. The bouncing threshold in this case can be 20 times higher than the threshold at the resonance. Similar features are observed in the case of Fano resonance in atomic physics \citep{Fano1961}. The maxima in the bouncing threshold correspond to antiresonance and will be the focus of this letter. \\

\begin{figure}[!]
	\centering
	\includegraphics[scale=.30]{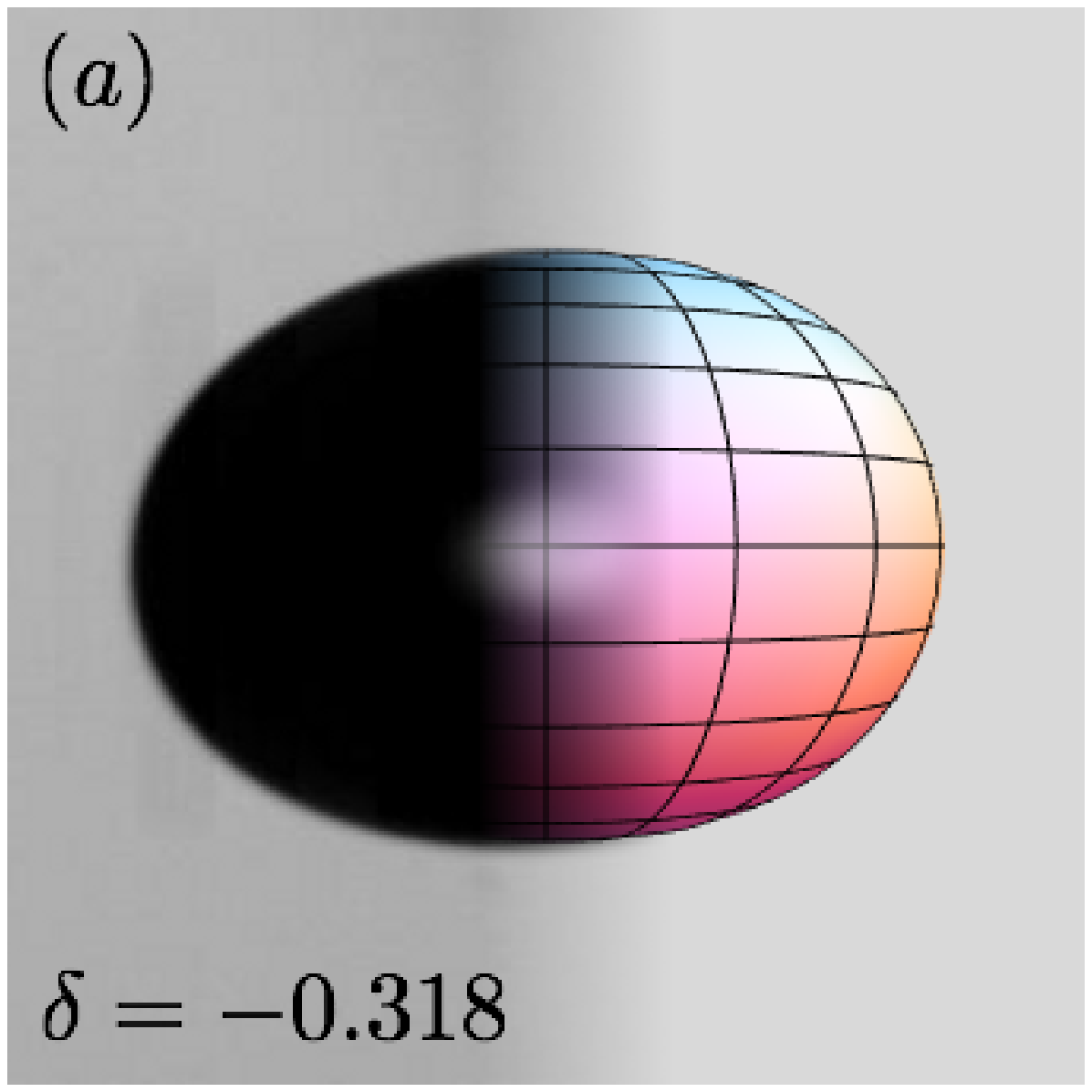} 
	\includegraphics[scale=.30]{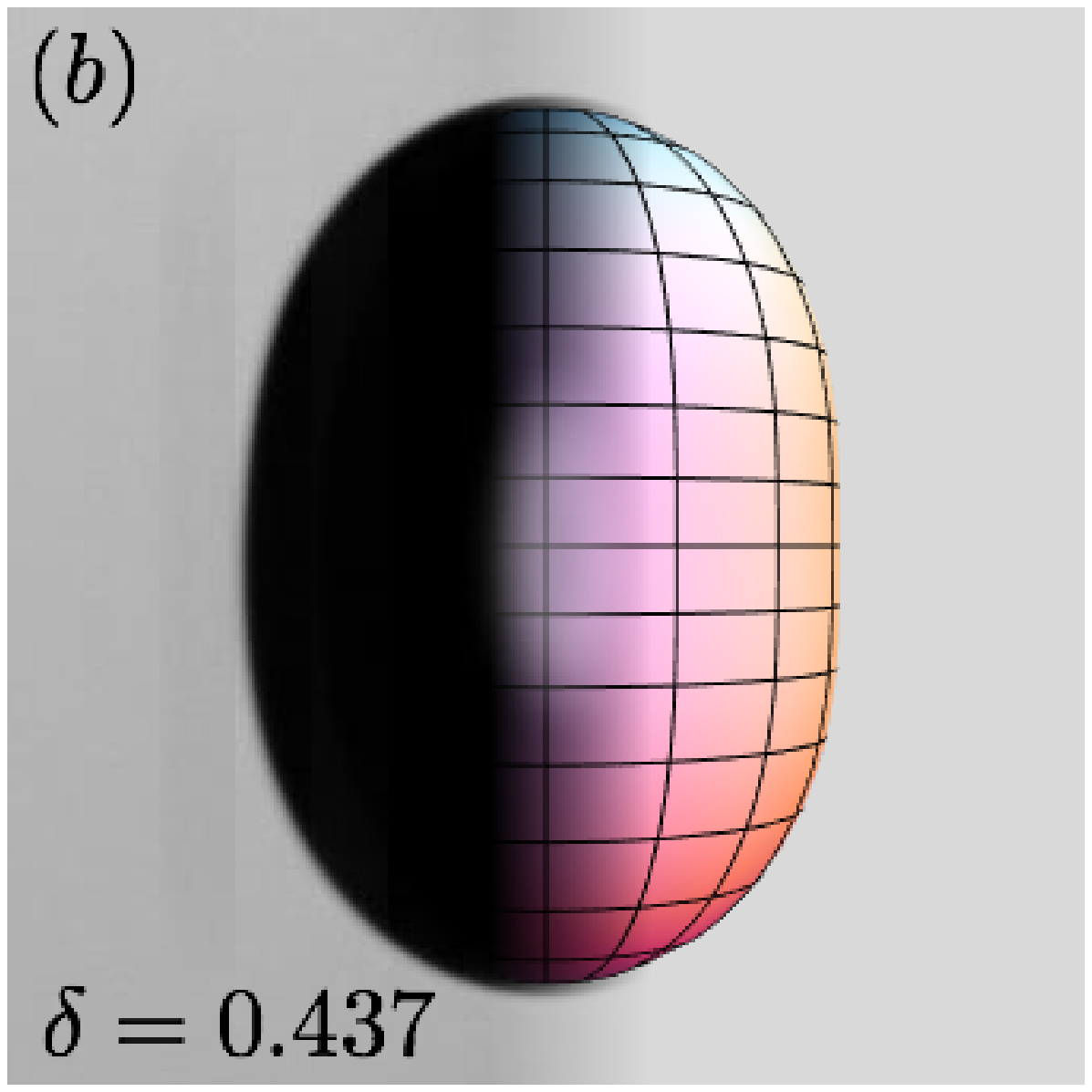} 
	\includegraphics[scale=.65]{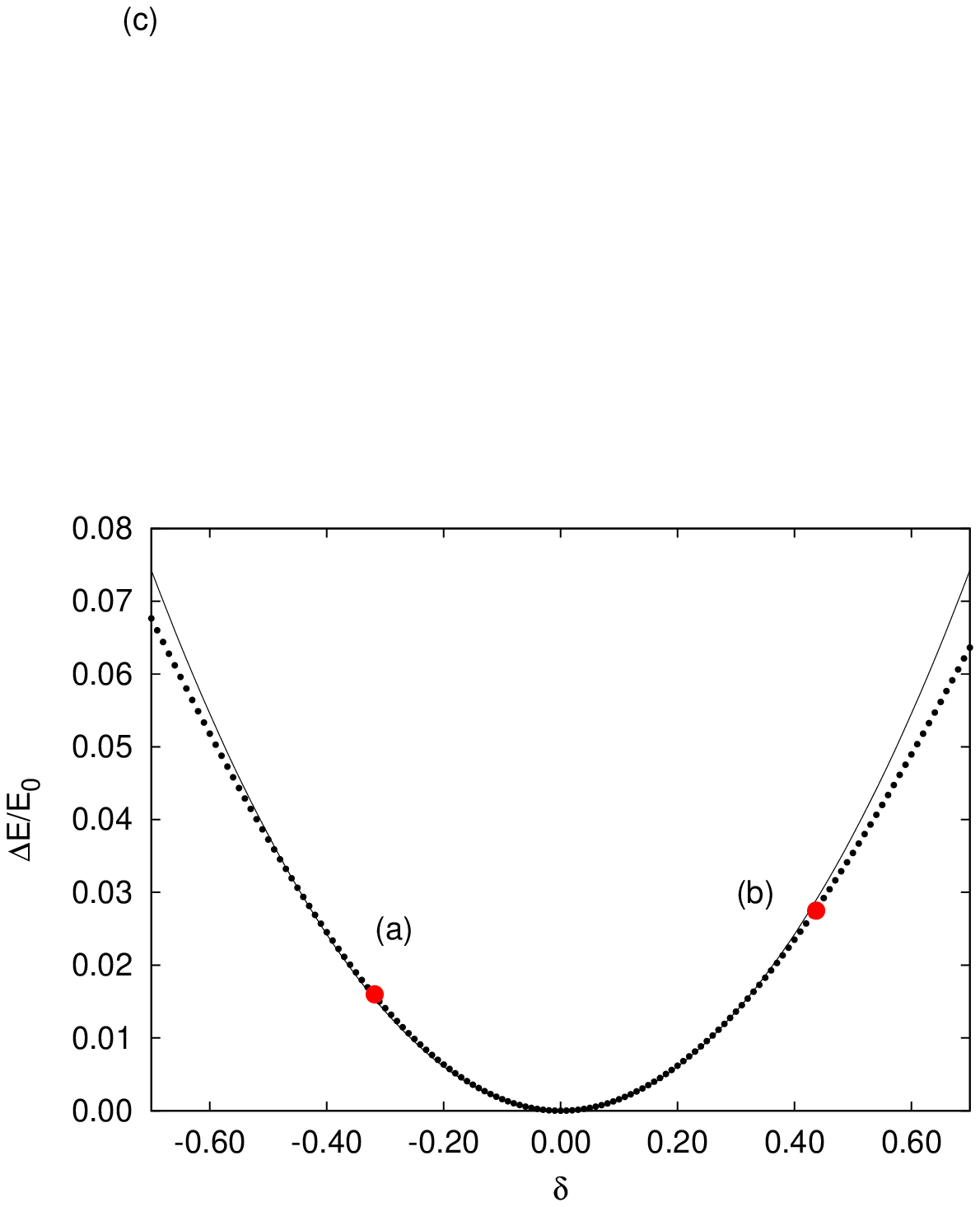} 
	\caption{(Color Online)(a) Droplet deformed with $\delta=-0.318$ resulting in an oblate shaped droplet. The left side is the experimental picture, the right one being the Prosperetti model.(b) Prolate shaped droplet with $\delta=0.437$. The value of $\delta$ in each case has been obtained by fitting the shape of the droplet.
	(c) Evolution of the relative surface energy for a droplet deformed with the $Y_2^0$ spherical harmonic, $E_0$ being the energy of an undeformed droplet (in black dots). The plain curve is a guide to the eyes showing the parabolic behavior. The x-axis measures the harmonics amplitude $\delta$. Note also the asymmetric shape of the curve.
	}{\label{fig:Schema}}
\end{figure}

In order to model BD, let us focus on the droplet shapes. The natural shape oscillations of a droplet have been describe by Prosperetti \citep{Prosperetti} with a series of spherical harmonics 
\begin{equation}
R(\theta,\phi)=a + \sum_{\ell=1}^{+\infty} \sum_{m=-\ell+1}^{\ell-1} c_{\ell} Y_\ell^m(\theta,\phi).
\end{equation}
The natural Rayleigh frequency $\omega_\ell$ \citep{Rayleigh} of each $\ell$ mode defines a dimensionless frequency $\Omega_\ell$ given by 
\begin{equation}\label{eqs:Rayleigh}
	\Omega_\ell=\omega/\omega_\ell=\sqrt{a^3\rho\omega^2/\sigma}\sqrt{1/\ell(\ell-1)(\ell+2)},
\end{equation}
where $\sigma$, $\rho$ and $a$ are respectively the droplet surface tension, density and radius. The parameter $\ell$ denotes the characteristic number of the considered spherical harmonic. Please note that because of the radius $a$ in the formula, the droplet can experience different value of $\Omega_l$ for a given set of parameters. This observation is the key ingredient of the droplet filter described here-below. By considering that the $Y^2_0$ axis-symmetric mode dominates others, as observed in our experiments (cf. Fig.\ref{fig:bounces}), one can write the droplet radius as the following decomposition
\begin{equation}
R(\theta,\phi)\approx a \left[ 1 +\delta Y_2^0(\theta,\phi) \right],
\end{equation}
where $\delta$ measures the maximum $Y_2^0$ deformation.  \\

Two images of a droplet closes to theoretical shapes are shown in Fig.\ref{fig:Schema}(a) and (b). They are characterized by oblate ($\delta=-0.318$) and prolate shapes ($\delta=0.437$). Computing the surface of the droplet under the appearance of the $Y_2^0$ spherical harmonic with the constraint of constant volume leads to the plot of Fig.\ref{fig:Schema}(c), showing the evolution of the relative surface energy $\Delta E / E_0$ with respect to $\delta$, $E_0$ being the surface energy of an undeformed droplet. One observes that $\Delta E / E_0 \propto \delta^2$. The droplet can therefore be considered as a linear spring. This observation is consistent with previous conclusions made in several articles \cite{Okumura2003,Biance2006,Dorbolo2008}.\\

\begin{figure}
	\centering
	\includegraphics[scale = 0.25]{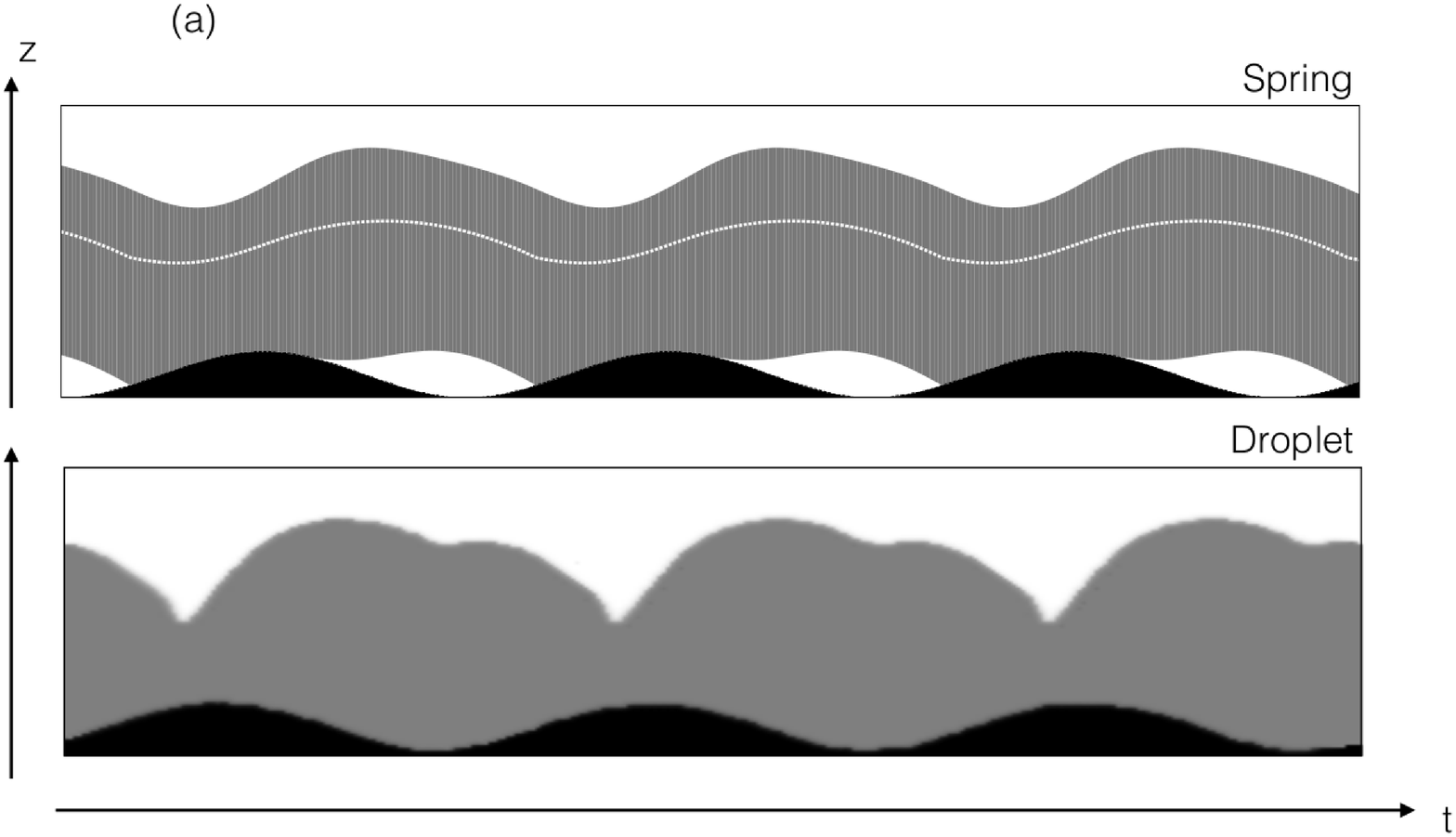} 
	\includegraphics[scale = 0.25]{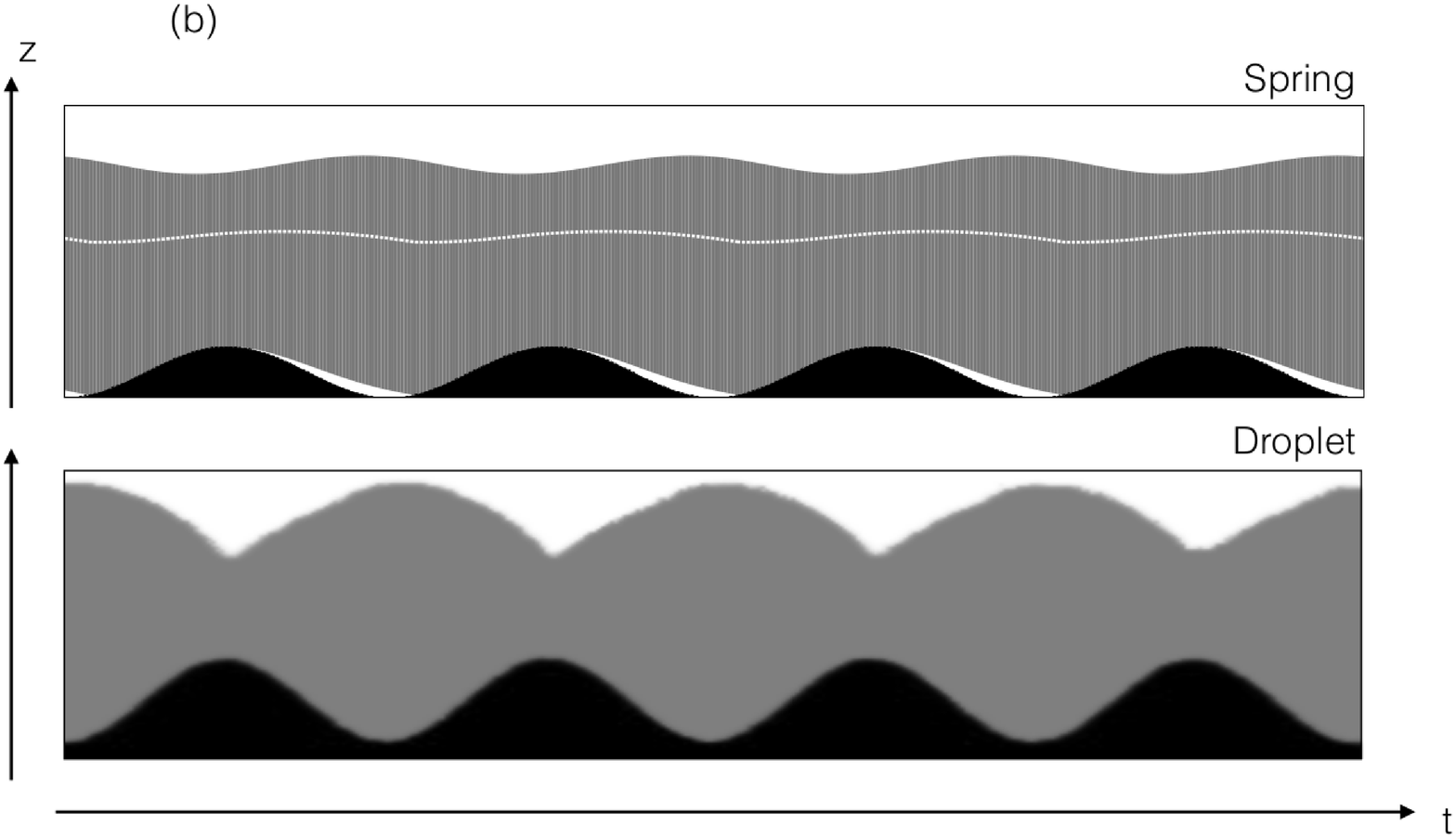}
	\caption{Numerical (noted ``Spring'') and experimental (noted ``Droplet'') spatio-temporal diagrams for BD with $\nu=5$ cSt (in arbitrary units). The white line in the spring motion represents its center of mass. The numerical parameters are the one indicated on Table \ref{tab:parameter}. (a) Resonant ABS/BD. (b) Antiresonant ABS/BD.}{\label{fig:Spatio}}
\end{figure}

Based on the above observations, one needs elasticity, damping and asymmetry in order to model the BD dynamics. Our model will focus on the axis-symmetric $Y^2_0$ harmonics. It considers two different masses $m_1$ and $m_2$ linked together by a spring (stiffness $k$ and natural length $L$) and by a damper (dissipation $\beta$). The masses $m_1$ and $m_2$ can be different in order to account for the asymmetric shapes observed during takes-off on Fig.1. The spring is used to give some stiffness in order to reproduce resonance and energy storage. The damper captures the dissipation within the droplet when it oscillates. The whole object bounces onto a rigid plate oscillating at the liquid surface amplitude $A$ and angular frequency $\omega$. The plate is chosen to be rigid since the bath beneath the droplet has a viscosity of 1000 cSt. Between two successive impacts, the ABS is only submitted to gravity $g$. Note that this model is unidimensional because of the $Y^0_2$ axial symmetry. Defining the mass distribution $\mu=m_1/(m_1+m_2)$, the spring natural frequency $\omega_0=\sqrt{k/(m_1+m_2)}$, the damper dissipation coefficient $\xi=\beta/2\omega_0(m_1+m_2)$ and  $\Omega_0=\omega/\omega_0$ the dimensionless oscillation frequency, $\Gamma=A\omega^2/g$ the dimensionless surface acceleration, $\phi=\omega t$ the dimensionless time, $l=L/A$ the dimensionless natural length and $\alpha=z/A$ the dimensionless height, Newton's second law of motion reads
\begin{equation}
\begin{cases}
 \alpha_p(\phi) =\cos(\phi),\\\label{eqs:equation}
 \ddot{\alpha}_1+\frac{2\xi \left(\dot{\alpha}_1-\dot{\alpha}_2\right)}{\Omega_0\mu}
+\frac{\left(\alpha_1-\alpha_2-l\right)}{\Omega^2_0\mu}+\frac{1}{\Gamma} =0,\\
 \ddot{\alpha}_2-\frac{2\xi \left(\dot{\alpha}_1-\dot{\alpha}_2\right)}{\Omega_0(1-\mu)}
-\frac{\left(\alpha_1-\alpha_2-l\right)}{\Omega^2_0(1-\mu)}+\frac{1}{\Gamma} =n_2(\phi).
\end{cases}
\end{equation}
The subscripts $p$, 1 and 2 are relative to the plate, the upper mass and the lower mass respectively. The dot above the symbols denotes the dimensionless time derivative, and $n_2$ is the dimensionless normal reaction. The ABS bouncing threshold can be obtained through those equations. Let us assume the following solutions for the motion of both masses:

\begin{align}
	&\alpha_1(\phi)=\alpha'\cos(\phi+\theta),\\
	&\alpha_2(\phi)=\cos(\phi).
\end{align}
After some algebra, the analytic expression of the bouncing threshold reads

\begin{equation}\label{eqs:Gamma}
	\Gamma_{th}(\Omega_0)=\sqrt{\frac{(1-\mu\Omega_0^2)^2+(2\xi\Omega_0)^2}{(1-(1-\mu)\mu\Omega^2_0)^2+(2\xi\Omega_0)^2}}.
\end{equation}
This bouncing threshold exhibits two extrema. Thus, the model is able to reproduce resonance and antiresonance. The minimum corresponds to resonance: the ABS stores some energy in its deformation and takes off for $\Gamma<1$. This dynamics is illustrated experimentally and numerically on Fig.\ref{fig:Spatio}(a). The maximum represents the antiresonance. The oscillation of the spring and the oscillation of the plate being phase shifted close to $\pi$, the spring center of mass feels a relative upward acceleration close to zero. As a consequence, the ABS takes off only for $\Gamma \gg 1$. The dynamics is illustrated on fig.\ref{fig:Spatio}(b). \\

In order to fit the experimental data, we forced the theoretical extrema to coincide with the experimental ones and came up with the plain curves on Fig.\ref{fig:Comp}. Previously proposed analytic models are also considered in this figure. The dotted curves correspond to the model proposed by Couder in \citep{Couder2005}. This model only considers the air layer beneath the droplet without any droplet nor bath deformation. The dashed curves are for the model proposed by Eichwald in \citep{Eichwald2010}. This model only considers the surface deformations. Couder's model is monotonically increasing and is thus unable to reproduce any of both extrema. Eichwald's model reproduces the minimum but fails in representing the maximum. The ABS model reproduces the general trend of the experimental data. As a conclusion, the antiresonant behavior is mainly due to the droplet deformations.  Beside those considerations, one could focus on the values of parameters $\xi$ and $\mu$ obtained through the fit. The fitting parameters are summed up on Table 1. As the droplet viscosity is increased, the value of both $\xi$ and $\mu$ increases. The behavior of $\xi$ can be rationalized if one remembers that this parameter models the dissipation within the droplet due to its deformations. Concerning $\mu$, each bounce leads to asymmetry as observed on Fig.\ref{fig:bounces}, and therefore $\mu\neq0.5$. Moreover, a larger viscosity implies a slower redistribution of mass in the bouncing object enhancing the asymmetry. Thus, $\mu$ increases with the droplet viscosity. \\

Resonance and antiresonance can be used to create a droplet size ``filter''. For this purpose, one has to consider that the size of a droplet is directly related to the natural Rayleigh frequency of the droplet (cf. Eq.(\ref{eqs:Rayleigh})). Since the antiresonance is found at this frequency and resonance at a specific lower frequency, the droplet « size selector » works as follows. The BD is driven at high acceleration $\Gamma$ much higher than the bouncing threshold $\Gamma_{th}$. For a given $\omega$, when the amplitude of vibration decreases, the acceleration could be lower than the antiresonant droplet condition. In that situation, all droplet sizes bounce except a specific one (band-stop filter). This is shown in the supplementary movie attached to this letter. While reducing again the amplitude of vibration, only the resonant droplets are able to bounce (band-pass fliter). A specific droplet size is therefore selected. We tested this procedure over a broad range of droplet sizes and we typically obtained a droplet diameter with a precision of 20 microns! This accurate technique could be exploited for improving all experiments \citep{Protiere2006,Dorbolo2008} for which droplet size is a dominant parameter.\\

This work was supported by the ULg ARC QUANDROPS and SUPERCOOL. MH thanks Yves-Eric Corbisier for his help concerning the videos. DR benefits a FRS-FNRS grant under the contract FRFC 2.4558.10. SD also thanks FNRS.

\end{document}